\documentclass[11pt, leqno]{article}
\usepackage{times, amsfonts, amsmath, amssymb, latexsym, setspace}
\usepackage{epsfig,subfigure, graphics}

\def\bs{\boldsymbol}
\def\Ex{{\rm I\!E}}
\def\Pr{{\rm I\!P}}
\def\be{\begin{equation}}
\def\ee{\end{equation}}
\def\bea{\begin{eqnarray*}}
\def\eea{\end{eqnarray*}}
\def\bean{\begin{eqnarray}}
\def\eean{\end{eqnarray}}
\def\nn{\nonumber}
\def\nin{\noindent}

\def\ra{\rightarrow}

\def\Bl{\Bigl}
\def\Br{\Bigr}

\def\R{{\bs{R}}}
\def\II{{\cal I}}
\def\JJ{{\cal J}}

\def\alp{\alpha}

\def\del{\delta}
\def\eps{\epsilon}
\def\lam{\lambda}

\def\Yn{\bs{Y}\!_n}
\def\Zn{\bs{Z}_n}
\def\Iapp{\bs{\II}_{app}}
\def\Ismall{\bs{\II}_{small}}
\def\normfn{\parallel \!\! f_n \!\! \parallel}

\newtheorem{Theorem}{Theorem}

\newtheorem{Lemma}{Lemma}

\begin{document}
\setstretch{1.5}

\title{Detection with the scan and the average likelihood ratio}
\author{Hock Peng Chan${}^{*}$ and Guenther Walther${}^{**}$ \\
        National University of Singapore and Stanford University}
\date{}
\maketitle

\begin{abstract}
We investigate the performance of the scan (the maximum likelihood
ratio statistic) and of the average likelihood ratio statistic
in the problem of detecting a deterministic signal with unknown
spatial extent in the prototypical univariate sampled data model
with white Gaussian noise. Our results show that the scan statistic,
a popular tool for detection problems, is optimal only for the 
detection of signals with the smallest spatial extent. For signals
with larger spatial extent the scan is suboptimal, and the power
loss can be considerable. In contrast, the average likelihood ratio
statistic is optimal for the detection of signals on all scales
except the smallest ones, where its performance is only slightly
suboptimal. We give rigorous mathematical statements of these results
 as well as heuristic explanations that suggest that the
essence of these findings applies to detection problems quite generally,
such as the detection of clusters in models involving densities
or intensities, or the detection of multivariate signals.
We present a modification of the average likelihood
ratio that yields optimal detection of signals with arbitrary
extent and which has the additional benefit of allowing for a
fast computation of the statistic. In contrast, optimal detection
with the scan seems to require the use of scale-dependent critical
values.
\end{abstract}

\vfill
\vfill

\noindent\textbf{Keywords and phrases.} Scan statistic, average likelihood
ratio statistic, optimal detection, fast algorithm.

\noindent\textbf{AMS 2000 subject classification.} 62G08, 62G10

\noindent${}^{*}$ Work supported by NUS grant R-155-000-090-112.\\
\noindent${}^{**}$ Work supported by NSF grant DMS-1007722 and NIH grant
AI077395.

\newpage

\section{Introduction and overview of results} \label{intro}

We are concerned with the problem of detecting a deterministic
signal with unknown spatial extent against a noisy background.
This problem arises in a wide range of applications, e.g. in
epidemiology and astronomy, and has received considerable attention
recently due to important problems in e.g. biosurveillance. The standard
statistical tool to address this problem is the scan statistic
(maximum likelihood ratio statistic), that considers the maximum
of local likelihood ratio statistics on certain subsets of the data.
There is a large body of work on scan statistics, see e.g. the
references in Glaz and Balakrishnan~(1999), Glaz, Naus, and Wallenstein~(2001), 
and Glaz, Poznyakov, and Wallenstein~(2009).
But there is also empirical evidence that the scan statistic is
suboptimal, see e.g. Neill~(2009) or Chan~(2009).

Siegmund~(2001) and Gangnon and Clayton~(2001) propose to use the 
average of the likelihood ratio
statistics instead of their maximum. In different contexts, various
versions of the average likelihood ratio where considered by Shiryaev~(1963),
Burnashev and Begmatov~(1990), and D\"{u}mbgen~(1998).
Chan~(2009) and Chan and Zhang~(2009) perform simulation studies for
various detection problems which suggest that the average likelihood ratio
statistic is superior to the scan statistic. In light of these
results, it is of interest to provide
a theoretical investigation of the performance of both the scan and the
average likelihood ratio. Such a theoretical comparison seems to be missing 
in the literature and appears to be quite relevant given the widespread use
of the scan statistic as a standard tool for a range of detection problems.

In the first part of this paper we show that
in the prototypical univariate sampled data
model with white Gaussian noise the
scan statistic possesses optimal detection power only for
signals with the smallest spatial extent; otherwise the scan
statistic is suboptimal, and the loss of power can be considerable
for signals having a large spatial extent. We also show that for
average likelihood ratio (ALR) statistic these conclusions hold in 
reversed order: The ALR possesses optimal detection power for
signals having large spatial extent, but is suboptimal for signals
with small spatial extent. However, the loss of power in the latter
case is so small that it is unlikely to be of concern, at least
for most sample sizes considered today.

In the second part of the paper we propose a modification of the ALR
that results in universal optimality and allows efficient computation.
The ALR averages the likelihood ratios pertaining to $\sim n^2$ 
stretches
of the data, where $n$ is the sample size, resulting in an $O(n^2)$
algorithm. Thus the use of the ALR is computationally infeasible
even for moderate sample sizes. We introduce a condensed ALR that
averages only a certain subset of the likelihood ratios and we show
that this condensed ALR possesses optimal detection power for signals
having arbitrary spatial extent. Furthermore, this condensed ALR can
 be computed in almost
linear time, viz. with an $O(n \log^2n)$ algorithm. In light of the
preceding discussion, it is arguably this improvement in computation time
rather than the small gain in detection power that is the main advantage
of this modification. We note that
typically, an approximation introduced to make a procedure computationally
less intensive will on the flip side degrade its performance somewhat. It is 
noteworthy that in the case of the ALR, our computationally efficient
modification actually leads to an improved (in fact: optimal)
performance.

We give sharp theoretical results on the performance of the ALR, the
scan statistic, and the newly proposed ALR in Sections~\ref{theory}
and ~\ref{modify}. Since 
these results are asymptotic, we complement them in Section~\ref{sims}  
with a simulation study that illustrates the results. 
Various modifications to the scan have been proposed in the literature
in order to improve its detection power.
We describe two such modifications in Section~\ref{modifiedscan},
and include them in our simulation study
to obtain a more informative comparison with the ALR.

As in the case of the ALR,
the computation of the scan statistic requires an $O(n^2)$ algorithm.
Various efficient
algorithms for computing a good approximation to the scan statistic
have been introduced in Neill and Moore~(2004), Arias-Castro, Donoho, and
Huo~(2005), Walther~(2010) and Rufibach and Walther~(2010).
Unlike the ALR, constructing a computationally efficient approximation
for the scan does not lead to universally optimal power.
Rather, statistical optimality for the scan seems to require the use of
size-dependent critical values.
We summarize our conclusions in Section~\ref{conclusion} and
defer proofs to Section~\ref{proofs}. 
The notation
$a_n \sim b_n$ means $c < a_n/b_n < C$ for constants
$0<c<C$.

\section{Comparison of the scan and the average likelihood ratio}   
\label{theory}

We observe
$$
Y_i\ =\ f_n\Bl(\frac{i}{n}\Br) +Z_i,\ \ \ \ i=1,\ldots,n,
$$
where the $Z_i$ are i.i.d. $N(0,1)$ and $f_n(x)=\mu_n {\bf 1}_{I_n}(x)$
with $I_n=(\frac{j_n}{n},\frac{k_n}{n}]$, $0\leq j_n<k_n\leq n$.
Both the amplitude $\mu_n$ and the support $I_n$ are unknown. The
task is to decide whether a signal is present, i.e. whether $\mu_n \neq 0$.

The above sampled data model with Gaussian white noise serves as
a prototype for many important applications. The heuristics and
results we develop below suggest that our conclusions carry over,
at least qualitatively, to related detection problems involving
multivariate signals, non-Gaussian errors, or the detection of clusters
in models involving densities or intensities, as described  
in Kulldorff~(1997).

The likelihood ratio statistic for testing $\mu_n=0$
when $I_n$ is known is computed as
$$
\exp \Bl(\frac{(\Yn(I_n))^2}{2}\Br),\ \ \ \mbox{ where } 
\Yn(I_n):=\frac{\sum_{i\in nI_n} Y_i}{\sqrt{n|I_n|}}=
\frac{\sum_{i=j_n+1}^{k_n} Y_i}{\sqrt{k_n-j_n}}.
$$

Since $I_n$ is unknown, the standard approach is to scan over
all intervals $I \in {\cal J}_n :=\{(\frac{j}{n},
\frac{k}{n}], 0\leq j<k\leq n\}$
for the largest likelihood ratio statistic. 
The resulting {\sl scan statistic} (maximum likelihood ratio
statistic) is 
$$
M_n\ :=\ \max_{0\leq j<k\leq n} \Bl|\Yn\Bl( \Bl(\frac{j}{n},
\frac{k}{n}\Br]\Br)\Br|.
$$
In contrast, the {\sl average likelihood ratio statistic} (ALR)
averages the likelihood ratios over all intervals $I \in {\cal J}_n$:
$$
A_n\ :=\ \frac{1}{n^2} \sum_{j=0}^n \sum_{k=j+1}^n
\exp \Bl(\frac{(\Yn( (\frac{j}{n},\frac{k}{n}]))^2}{2}\Br).
$$

To quantify the performance of these statistics, we look for the smallest
value of $|\mu_n|$ that allows a reliable detection of the signal.
As explained below, in order to achieve
optimality a test must be able to asymptotically detect signals
$f_n$ with 
\be  \label{optcondition}
|\mu_n|\sqrt{|I_n|} \ \geq \ \frac{\sqrt{2 \log \frac{1}{|I_n|}}
+b_n}{\sqrt{n}},\ \ \ \mbox{ where } b_n \ra \infty.
\ee
Note that for signals $f_n$ on {\sl small scales}, 
$|I_n| \ra 0$, (\ref{optcondition}) is equivalent to
\be  \label{optsmall}
|\mu_n|\sqrt{|I_n|} \ \geq \ (\sqrt{2} +\eps_n)
\sqrt{ \frac{\log \frac{1}{|I_n|}}{n}},
\ee
where $\eps_n$ can go to 0 but not too fast: $\eps_n \sqrt{
\log \frac{1}{|I_n|}} \rightarrow \infty$. 

For signals $f_n$ on {\sl large scales}, $ \liminf_n
|I_n| > 0$, (\ref{optcondition}) is equivalent to
\be  \label{optlarge}
|\mu_n| \ \geq \ \frac{b_n}{\sqrt{n}},
\ \ \ \mbox{ where } b_n \ra \infty.
\ee
It is impossible to detect signals with noticeably smaller mean:
In the case of signals on small scales, a classical argument
in the minimax framework
(see e.g. Lepski and Tsybakov~(2000), D\"{u}mbgen and Spokoiny~(2001),
and D\"{u}mbgen and Walther~(2008)) shows that if `$+\eps_n$' is replaced
by `$-\eps_n$' in (\ref{optsmall}), then there exists no test that
can detect such $f_n$ with nontrivial asymptotic power. Likewise,
a contiguity argument, as in D\"{u}mbgen and Walther~(2008), shows
that in the case of large scales the condition (\ref{optlarge}) is
necessary for any test to be consistent against $f_n$. 
On the other hand, we exhibit below a test that detects 
signals satisfying (\ref{optcondition}) with
asymptotic power 1.
Thus the detection threshold given by (\ref{optcondition})
marks a standard that
is attainable but cannot be improved upon. We now examine how
the scan and the ALR compare against this standard.

\begin{Theorem}  \label{scan}
Let $\kappa_n$ be the $(1-\alp)$ quantile of the null distribution of $M_n$.
\begin{enumerate}
\item If $|\mu_n|\sqrt{|I_n|} \geq (\sqrt{2}+\eps_n)\sqrt{\frac{\log n}{n}}$
with $\eps_n \sqrt{\log n} \rightarrow \infty$,
then $\Pr_{f_n} (M_n > \kappa_n) \rightarrow 1$.
\item If $|\mu_n|\sqrt{|I_n|}=(\sqrt{2}-\eps_n)\sqrt{\frac{\log n}{n}}$
with $\eps_n$ as above,
then $\overline{\lim}_n \Pr_{f_n}(M_n > \kappa_n) \leq \alp$.
\end{enumerate}
\end{Theorem}

Thus the detection threshold for the scan is $\sqrt{2\frac{\log n}{n}}$,
irrespective of the spatial extent of the signal. 
Comparing to (\ref{optsmall}), one sees that
the scan is optimal only for signals having the smallest spatial
extent, i.e. for $|I_n|$ close to $\frac{1}{n}$. As an illustration,
if $|I_n|=n^{-p}$, $p \in (0,1]$, then detection is possible only
if $|\mu_n|\sqrt{|I_n|}$ is at least $p^{-1/2}$ times larger than the optimal
threshold. In the case of large scales, comparison
with (\ref{optlarge}) shows that this multiplier diverges
to infinity, and thus the scan suffers from a noticeably inferior
performance. These results are illustrated in the simulation study
in Section~\ref{sims}, and explain the sometimes disappointing
performance of the scan observed in the literature.

We note that an alternative way to analyze the performance
of the scan is to put a prior on the unknown spatial extent of
the signal, e.g. the uniform distribution on $(0,1)$. It is
readily seen that this analysis leads to the same conclusions
as the case of large scales above, i.e. the scan is far from
optimal.

The next theorem details the performance of the average likelihood
ratio:

\begin{Theorem}  \label{ALR}
Let $\tau_n$ be the $(1-\alp)$ quantile of the null distribution of $A_n$.
\begin{enumerate}
\item $A_n$ is optimal for detecting signals with large spatial extent:\\
If $ \liminf_n |I_n| > 0$ and $|\mu_n| = \frac{b_n}{\sqrt{n}}$ with
$b_n \ra \infty$, then $\Pr_{f_n} (A_n > \tau_n) \rightarrow 1$.
\item $A_n$ is not optimal for detecting signals with small spatial 
extent:\\
If $|I_n| \ra 0$ and $|\mu_n|\sqrt{|I_n|} =K\sqrt{ \frac{\log \frac{1}{|I_n|}
}{n}}$
with $K<2$, then $\overline{\lim}_n \Pr_{f_n}(A_n > \tau_n) \leq 
\alp$.
\item If $K \geq 2+\eps_n$, where $\eps_n \sqrt{
\log \frac{1}{|I_n|}} \rightarrow \infty$, then
$\Pr_{f_n} (A_n > \tau_n) \rightarrow 1$.
\end{enumerate}
\end{Theorem}

Comparing with (\ref{optsmall}), one sees that on small scales the ALR
requires $|\mu_n|\sqrt{|I_n|}$ to be about $\sqrt{2}$ times larger than the optimal  
threshold. This discrepancy
is not very consequential: The simulations in Section~\ref{sims} show 
that the corresponding
loss of power is quite small for sample sizes up to $n=10000$,
which is the largest sample size we were able to simulate due to
the $O(n^2)$ computational complexity of the ALR.

A heuristic explanation of why the scan and the ALR do not obtain 
optimality is as follows: There are $n$ disjoint intervals $I$ of
length $1/n$. The corresponding likelihood ratio statistics $\Yn(I)$
are i.i.d. N(0,1) under the null hypothesis, thus their maximum
behaves like $\sqrt{2 \log n}$.
But in the case of large intervals of length $1/c$ (say), there are
only $c$ disjoint intervals that result in independent statistics
$\Yn(I)$. The statistics for the other intervals of length $1/c$ 
are not independent of these $\Yn(I)$ since the intervals overlap. 
Thus the null distribution of that maximum behaves roughly like
the maximum of $c$ i.i.d. N(0,1), which is $O_p(1)$. Hence the
overall maximum $M_n$ is dominated by the small intervals, with
a corresponding loss of power at large intervals. 

As for the ALR, if a detectable signal lives on a large interval $I_n$,
then $\Yn(I)$ is significant provided $I$ has a nonvanishing
overlap with $I_n$. Since there are $\sim n^2$ such intervals, the
ALR is significant despite the divisor $n^2$ in its definition.
In the case of small intervals $I_n$, however, the number of intervals
$I$ that yield a sufficiently large statistic $\Yn(I)$ is so small
compared to the total number of intervals ($\sim n^2$) that their
contribution to $A_n$ is annihilated by the divisor $n^2$. More precisely:
the likelihood ratio statistic is maximized at $I=I_n$, where its size
is $\gg |I_n|^{-1}$ (up to log terms) for signals at the detection
threshold (\ref{optcondition}). Thus if $|I_n|=1/n$, then there are only
a few significant likelihood ratios and their magnitude is 
about $|I_n|^{-1}=n$. Thus dividing by $n^2$ will let their contribution 
vanish unless the size of the likelihood ratio statistics is
increased to $|I_n|^{-2}=n^2$ by doubling $|\mu_n|^2|I_n|$ in the
log likelihood ratio.

\section{The condensed average likelihood ratio statistic}
 \label{modify}

The above heuristic suggests that an optimal version of the
ALR can be constructed by averaging the likelihood ratios not
over all $\sim n^2$ intervals of ${\cal J}_n$ but over a subset
of ${\cal J}_n$ with cardinality close to $n$. The general idea
is that for larger
intervals, there is not much lost by considering only intervals with
endpoints on a coarser grid as long as the distance between such gridpoints
is small compared to the length of the intervals. Then these
intervals still provide a good approximation to ${\cal J}_n$,
while the cardinality of this approximating set can be reduced
dramatically. To implement this idea, we modify the approach in 
Walther~(2010) and Rufibach and Walther~(2010) and
group intervals into $\ell_{max}=\lceil \log_2 \frac{n}{\log n}\rceil$
sets, each of which contains intervals having about the same length:
the approximating set 
$\Iapp(\ell)$ consists of intervals that contain between $m_{\ell}+1$
and $2m_{\ell}$ design points and whose endpoints are restricted to
a grid consisting of every $d_{\ell}$th design point, where 
$m_{\ell}=n2^{-\ell}$
and $d_{\ell}=\Bl\lceil \frac{\sqrt{m_{\ell}}\ell^{4/5}}{\log n}
\Br\rceil$. Our overall approximating set is then the union of these
$\Iapp(\ell)$ together with all small intervals:
\begin{align}
\Iapp &= \bigcup_{\ell =1}^{\ell_{max}} \Iapp(\ell) \ \ \cup \ \ \Ismall,
\ \ \mbox{ where } \nn \\
\Iapp(\ell) &= \Bl\{(\frac{j}{n},\frac{k}{n}] \in {\cal J}_n: 
j,k \in \{i d_{\ell},
i=0,1,\ldots \}\ \mbox{ and } m_{\ell}<k-j\leq 2m_{\ell}\Br\},\nn \\
\Ismall &= \Bl\{(\frac{j}{n},\frac{k}{n}]\in {\cal J}_n:
k-j \leq m_{\ell_{max}}\Br\}.\nn
\end{align}
We suppress the dependence on $n$ for notational simplicity.
Our {\sl condensed ALR} is thus
$$
A_{n,cond}\ :=\ \frac{1}{\# \Iapp} \sum_{I \in \Iapp}
\exp \Bl(\frac{(\Yn( I))^2}{2}\Br).
$$
The above choices of $d_{\ell}$ and $m_{\ell}$ result in statistical and
computational efficiency for the ALR and differ from the choices used
in Walther~(2010) and Rufibach and Walther~(2010).
We give an explanation for this  in
the proof of Theorem~\ref{ALRapprox}. 

\begin{Theorem}  \label{ALRapprox}
The condensed ALR $A_{n,cond}$ is optimal for detecting signals with arbitrary
 spatial extent. Furthermore,
$A_{n,cond}$ can be computed in $O(n \log^2\!n)$ time.
\end{Theorem}

\section{Modifications of the scan that improve power}  \label{modifiedscan}

Here we describe two simple ways to improve the power of the scan
by fixing the miscalibration across different scales described in
Section~\ref{theory}. For the first one we adapt the
penalty term introduced by D\"{u}mbgen and Spokoiny~(2001)
in the context of inference about a function.
The idea is to subtract off
the putative maximum at each scale in order to put the different
scales on an equal footing: the {\sl penalized scan} is
$$
P_n\ :=\ \max_{0\leq j<k\leq n} \Bl(\Bl|\Yn\Bl( \Bl(\frac{j}{n},
\frac{k}{n}\Br]\Br)\Br|-\sqrt{2 \log \frac{en}{k-j}}\Br).
$$
We declare that a signal is present if $P_n > \gamma_n(\alp)$,
where $\gamma_n(\alp)$ is the $(1-\alp)$ quantile of the null
distribution of $P_n$.

A drawback of the penalized scan is that it requires
the specification of the penalty term, which
has to be derived for each situation at hand. The 
penalty term $\sqrt{2 \log \frac{en}{k-j}}$ optimizes signal
detection for all scales in the Gaussian regression setting;
a different setting or a different error distribution
may require a different penalty term to achieve optimal detection.
The form
of the penalty term depends on the tail behavior of the
local test statistics, their dependence structure, and the entropy
of the underlying space, see Theorem~7.1 in D\"{u}mbgen and Walther
(2008). Thus these properties
have to be derived on a case-by-case basis, and this derivation is
typically far from straightforward.

The second way to fix the miscalibration of the scan is the
{\sl blocked scan} introduced in Walther (2010).
The block method has the advantage that it is a general recipe
that does not require any case-specific input.
The idea is to group intervals having roughly the same length into
blocks, with the $\ell$th block comprising all intervals that
contain between $m_{\ell}=n2^{-\ell}$ and $2m_{\ell}$ design
points. Then one assigns different critical values to different
blocks such that the significance level on the $\ell$th block
decreases as $\sim \ell^{-2}$.

In more detail, for $m_{\ell}$ and $\ell_{max}$ as above, define
$$
M_{n,\ell}\ :=\ \max_{m_{\ell}<k-j\leq m_{\ell-1}} \Bl|\Yn\Bl( \Bl(
\frac{j}{n},\frac{k}{n}\Br]\Br)\Br|,\ \ \ \ \ \ell=1,\ldots,\ell_{max},
$$
and $M_{n,\ell_{max}+1}:=\max_{k-j\leq m_{\ell_{max}}}
|\Yn( (\frac{j}{n},\frac{k}{n}])|$. Thus the
($\ell_{max}+1$)st block comprises all small intervals that contain up
to $m_{\ell_{max}}$ design points. The {\sl blocked scan} declares
that a signal is present if $M_{n,\ell} > q_{\ell}\Bl(\frac{\tilde{\alp}}{
(A+\ell)^2} \Br)$ for any $\ell \in \{1,\ldots,\ell_{max}+1\}$.
Here $q_{\ell}\Bl(\frac{\tilde{\alp}}{(A+\ell)^2} \Br)$ is the
$\Bl(1-\frac{\tilde{\alp}}{(A+\ell)^2} \Br)$ quantile of the null
distribution of $M_{n,\ell}$, $A:=10$ (say), and $\tilde{\alp}$ is
chosen such that the overall significance level is $\alp$:
$$
\Pr_0 \Bl( \bigcup_{\ell=1}^{\ell_{max}+1}\Bl\{ M_{n,\ell} >
q_{\ell}\Bl(\frac{\tilde{\alp}}{(A+\ell)^2} \Br) \Br\}\Br) \ =\ \alp.
$$
The critical values $q_{\ell}$ and $\tilde{\alp}$ can be easily
simulated with Monte Carlo, see Rufibach and Walther (2010). We
suppress the dependence of $q_{\ell}$ on $n$ for notational
simplicity.

It can be shown that both the penalized scan and the blocked
scan are also optimal for detecting signals with arbitrary spatial
extent, see Chan and Walther~(2011).

Computationally efficient algorithms for evaluating the scan
or an approximation thereof have been introduced in the
literature, see e.g. Neill and Moore~(2004), Arias-Castro
et al.~(2005), Rufibach and Walther~(2010) and Walther~(2010).
Those algorithms reduce the computational complexity for the scan from $O(n^2)$
to almost linear time in $n$ (apart from $\log n$ factors), 
comparable to
the condensed scan.
Hence it is possible to modify the scan to obtain statistical optimality
(via the penalized or blocked scan) and computational efficiency.
But, unlike the case of the condensed ALR where the particular choice of the
approximating set leads to optimal power properties, it appears
that evaluating the scan on an appropriate approximating set does not lead
to optimal detection by itself. Rather, it appears that optimal
detection requires the use of scale-dependent critical values, and efficient
computation has to be addressed separately using any of the methods
cited above.

\section{A simulation study}  \label{sims}

\begin{figure}[h]
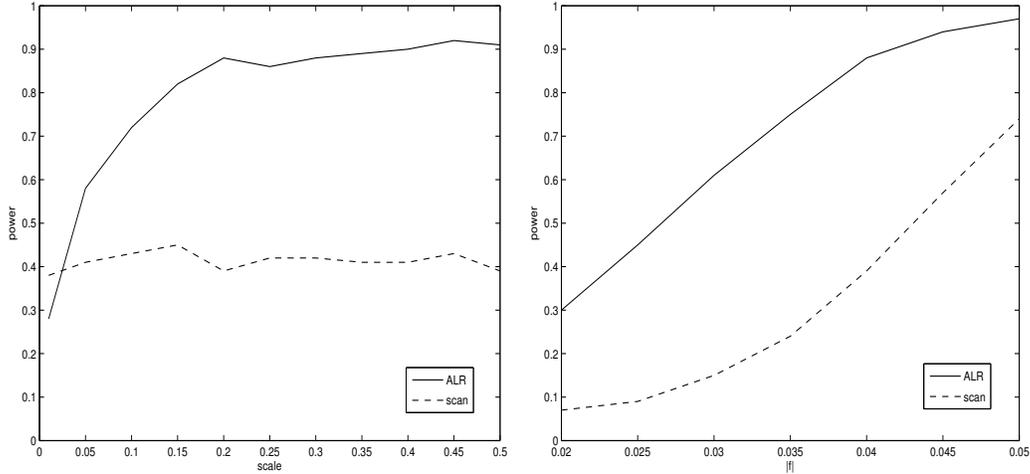

\centering
\subfigure{
 \includegraphics[width=0.45\textwidth,height=0.28\textheight,
clip]{powerplot.pdf}}
\subfigure{
 \includegraphics[width=0.45\textwidth,height=0.28\textheight,
clip]{powerplot2.pdf}}\\
\caption{Left: Power of the scan and the ALR for detecting signals $f_n$
with fixed norm $\normfn=0.04$ but varying spatial extent $|I_n|$, $n=10000$.
Right: Power of the scan and the ALR for detecting signals $f_n$ with
varying norms $\normfn$ and random spatial extent. 
The power curves for the condensed ALR,
the penalized scan, and the blocked scan are similar to those of the ALR
and are not plotted, see Tables~\ref{table1} and~\ref{table2} for
the numerical results.}
\label{fig1}
\end{figure}

Since the results of the previous sections are asymptotic,
we illustrate them in a finite
sample context with a simulation study.
We first consider signals $f_n$ with fixed norm $\normfn :=|\mu_n| 
\sqrt{|I_n|}$ but varying
spatial extent. Table~\ref{table1} gives the power of the scan, the ALR,
the condensed ALR, the penalized scan, and the blocked scan for a sample
size of $n=10000$. The results are visualized 
in the left plot in Figure~\ref{fig1}.
One sees that the overall performance of the scan is inferior to that
of the other four methods, whose performances are quite similar.
In particular, the  power of the scan is not increasing with the spatial
extent of the signal as opposed to the other four methods.
As a consequence, the scan is competitive only for signals
on the smallest scales. 

Table~\ref{table1} also shows an 
improvement in power of the condensed ALR vis-a-vis the ALR on small
scales, illustrating Theorems~\ref{ALR} and ~\ref{ALRapprox}. However, this
improvement is modest, at least for the sample size under consideration,
and thus the main advantage of the
condensed scan is arguably the dramatic reduction in computation
time to $O(n \log^2n)$ versus $O(n^2)$ for the ALR. We were
able to accurately simulate critical values for the condensed
ALR with a sample size of 1 million in a matter of hours, whereas
this computation would take hundreds of days for the ALR.

Table~\ref{table2} shows how the power varies as function of $\normfn$,
see the right plot in Figure~\ref{fig1} for a visual representation.
The spatial extent of the signal was chosen uniformly in $[0,1]$ in each
of the 2000 Monte Carlo simulations. The power curves of the last four
methods are again quite similar, and superior to that of the scan.
One sees that the scan requires a signal with almost twice the norm
to achieve the power of the four other methods. According to the
results in the previous sections, this discrepancy increases with
the sample size.

\begin{table}[h]
\begin{tabular}{r|ccccccccccc}
scale & 0.01& 0.05& 0.1& 0.15& 0.2 &  0.25 & 0.3 & 0.35 & 0.4 & 0.45 & 0.5\\
\hline
scan &  38 & 41 & 43 & 45 & 39 & 42 & 42 & 41 & 41 & 43 & 39\\
ALR&    28 & 58 & 72 & 82 & 88 & 86 & 88 & 89 & 90 & 92 & 91\\
condensed ALR& 36 & 61 & 72 & 80 & 87 & 85 & 87 & 88 & 90 & 91 & 91\\
penalized scan& 37 & 61 & 72 & 80 & 85 & 84 & 85 & 86 & 87 & 90 & 89\\
blocked scan&  41  &59  &69 & 77 & 82 & 80 & 82 & 82 & 84 & 87 & 86\\
\end{tabular}
\caption{ Power in percent for detecting signals $f_n$ with fixed norm
$\normfn=0.04$ but varying spatial extent $|I_n|$, $n=10000$.}
\label{table1}
\end{table}

\bigskip

\begin{table}[h]
\begin{tabular}{r|ccccccc}
$\normfn \times 100$ & 2& 2.5& 3& 3.5& 4 &  4.5 & 5 \\
\hline
scan &  7 & 9 & 15 & 24 & 39 & 57 & 74 \\
ALR&    30 & 45 & 61 & 75 & 88 & 94 & 97\\
condensed ALR& 30  &44 & 60 & 75 & 87 & 94 & 97\\
penalized scan& 26&  40&  57&  74&  85&  93&  97\\
blocked scan&  24  &35 & 51&  69&  82&  92&  96\\
\end{tabular}
\caption{ Power in percent for detecting signals $f_n$ with 
varying norms $\normfn$ and random spatial extent, $n=10000$.}
\label{table2}
\end{table}

All power values in Tables~\ref{table1} and \ref{table2} are
with respect to a 5\% significance level. The corresponding
critical values were simulated with 10000 Monte Carlo samples,
and the power was simulated with 
2000 Monte Carlo samples. The location of the signal was chosen
at random in each of these simulations to avoid confounding the
results with the approximation scheme of the condensed ALR.

\section{Conclusion}  \label{conclusion}

The scan is optimal only for detecting signals on the smallest
scales. The ALR has a superior overall performance and is optimal for 
detecting signals on all scales except on the smallest ones, but
the loss of power there appears to be modest. Moreover, by averaging
the likelihood ratios over a particular subset of intervals rather
than over all intervals, the resulting condensed ALR is simultaneously
optimal for all scales and also allows for efficient computation.
In contrast, improved versions of the scan, such as the penalized
scan and the blocked scan, appear to require the use of scale-dependent
critical values, and thus it appears that statistical optimality and
computational efficiency have to be addressed separately for the scan.

The results of this paper are developed in the  Gaussian
white noise model since it is known that the conceptual results in that model
are applicable and relevant for a wide range of related problems.
We note that the concrete implementation of the results derived
in the Gaussian white noise model requires additional work that
depends on the concrete problem at hand. For example, in the univariate
regression setting, Rohde~(2008) employs local signed rank tests to transform
non-Gaussian data into statistics with sub-gaussian tails, and Cai, Jeng, and
Li~(2011)
employ a local median transformation for the same purpose. To see why such
an additional step is required, note that the null distribution of both the 
scan and the ALR, as well as the form of the penalty term for the penalized
scan, depend sensitively on the tails of the error distribution, and hence 
on the assumption of Gaussianity. The above papers show rigorously
that the Gaussian white noise model is applicable after a local signed rank
or a local median transformation, assuming only e.g. symmetry of the error 
distribution. These arguments are technically sophisticated and thus the
transformation step constitutes a piece of methodological work by itself.
It is thus helpful 
to separate the conceptual issues involving the scan and the ALR from the
particular implementation and to present an unencumbered exposition
in the Gaussian white noise model.
In particular, the heuristics developed in this model give
guidance how the corresponding problems might be addressed in related detection
problems, such as the detection of clusters in models
involving densities and intensities, as well as the important case of 
detecting multivariate signals. For example, in a regression type setting
with irregularly
spaced or multivariate data, the proportion of observations falling
into the set $I$ would take the place of the size $|I|$ used above.
Section~\ref{modify}   shows how to construct an approximating set that, on
the one hand, results in computational efficiency and, on the other, 
provides the right `weighting' of the various sizes of intervals in the
condensed ALR to achieve statistical optimality. This idea can
presumably be mimicked in a multivariate situation, where the construction
would then depend on the entropy of the class of scanning windows.
The implementation of this idea in the multivariate context is
an interesting problem for future research. Another open problem
is a theoretical result on the precision with
which the scan and the ALR allow one to localize a signal once it is detected.

\section{Proofs}  \label{proofs}

Note that $f_n(x)=\mu_n {\bf 1}_{I_n}(x)$ implies for any interval
$I \in {\cal J}_n$:
\be \label{1a}
\Yn(I)\ =\ \Zn(I)+ sign(\mu_n)\sqrt{n} |\mu_n| \frac{|I \cap I_n|}{
\sqrt{|I|}}.
\ee

We use the following consequence of a result of D\"{u}mbgen and 
Spokoiny~(2001).
\begin{Lemma}  \label{B1}
Let $\ell \in (0,1)$ and $J \in {\cal J}_n$, where $J$ does not depend 
on $\Zn$. Then
$$
\max_{I \in  {\cal J}_n:I \subset J, |I|\geq \ell} |\Zn(I)| \ 
\stackrel{d}{\leq} \ L+\sqrt{2 \log \frac{e|J|}{\ell}}
$$
for a universal random variable $L$ which is finite almost surely.
\end{Lemma}
\bigskip

{\bf Proof of Lemma~\ref{B1}:} Writing $W$ for Brownian motion and $j,k$
for integer indices:
\begin{equation} \label{B0}
\begin{split}
\max_{\substack{
     I \in {\cal J}_n: I \subset J \\
     |I| \geq \ell
     }}  \Bl(|\Zn(I)|-\sqrt{2 \log \frac{e|J|}{|I|}}\Br)
& \stackrel{d}{=} \max_{\substack{
                       0\leq j<k\leq n|J|\\
                       k-j \geq n\ell}}
\Bl(\frac{|W(k)-W(j)|}{\sqrt{k-j}} -\sqrt{2 \log \frac{e|J|n}{k-j}}\Br)\\
& \leq \sup_{\substack{
             s,t \in \R:\\
             0\leq s<t \leq n|J|}}
\Bl(\frac{|W(t)-W(s)|}{\sqrt{t-s}} -\sqrt{2 \log \frac{e|J|n}{t-s}}\Br)\\
& \stackrel{d}{=} \sup_{0 \leq s<t\leq 1}
\Bl(\frac{|W(t)-W(s)|}{\sqrt{t-s}} -\sqrt{2 \log \frac{e}{t-s}}\Br)
\ =:\ L 
\end{split}
\end{equation}
by Brownian scaling. Thus the random variable $L$ defined above is 
universally applicable for all $n, \ell$, and $J$. Importantly, $L$
is finite almost surely, see Sec. 6.1 in D\"{u}mbgen and Spokoiny~(2001).
\hfill $\Box$
\bigskip

{\bf Proof of Theorem~\ref{scan}:}
As for part~1, (\ref{1a}) implies 
$M_n \ \geq \ |\Yn(I_n)| \ \geq \ -|\Zn(I_n)|+\sqrt{2 \log n} 
+\eps_n \sqrt{\log n}$.
Since $\Zn(I_n) \sim N(0,1)$ and $\eps_n \sqrt{\log n} \rightarrow \infty$,
the claim follows from $\kappa_n=\sqrt{2 \log n}+O(1)$, see (\ref{kappa}).

For the proof of part~2, set $b_n:=\eps_n \sqrt{\log n} \ra \infty$
and consider first the collection of intervals ${\cal J}_{n,1}:=
\Bl\{I \in {\cal J}_n:\ I \cap I_n \neq \emptyset$ and $|I_n|/b_n \leq 
|I| \leq b_n |I_n|\Br\}$. So $I \in {\cal J}_{n,1}$ implies
$I \subset \Bl(\frac{j_n -\lfloor b_n n |I_n|\rfloor}{n},
\frac{k_n+\lfloor b_n n |I_n|\rfloor}{n} \Br] \cap (0,1]$
and thus Lemma~\ref{B1} gives 
$$
\max_{I\in {\cal J}_{n,1}}|\Zn(I)|\ \
\stackrel{d}{\leq} \ \ L+\sqrt{2 \log \frac{e(1+2b_n)|I_n|}{|I_n|/b_n}}
\ \ \leq \ \ L+2\sqrt{\log (3b_n)}.
$$
Together with (\ref{1a}) and
$|I \cap I_n| \leq \sqrt{|I||I_n|}$, this yields
\begin{equation*}
\begin{split}
\Pr_{f_n}\Bigl( \max_{I\in {\cal J}_{n,1}}|\Yn(I)| > \kappa_n \Br)
& \leq \Pr\Bigl( \max_{I\in {\cal J}_{n,1}}|\Zn(I)| +\sqrt{2 \log n}
 -b_n > \kappa_n \Br) \\
& \leq \Pr\Bl( L > \kappa_n -\sqrt{2 \log n} +b_n-2\sqrt{\log (3b_n)} \Br)\\
& \ra 0\ \ \ \mbox{ since } \kappa_n =\sqrt{2 \log n} +O(1) 
  \mbox{ by (\ref{kappa}).}
\end{split}
\end{equation*}

Next, only if $b_n \leq \log^3\!n$ do we need to consider
${\cal J}_{n,2}:=\Bl\{I \in {\cal J}_n:\ I \cap I_n \neq \emptyset$ and
either $|I_n|/\log^3\!n \leq |I| \leq |I_n|/b_n\ $ or $\ b_n|I_n| <|I| \leq
|I_n| \log^3\!n\Br\}$. Similarly as above, $I \in {\cal J}_{n,2}$ implies
that $I$ is contained in an interval $J \in {\cal J}_{n}$ with
$|J|\leq (1+2\log^3\!n)|I_n|$. Thus Lemma~\ref{B1} yields 
$$
\max_{I\in {\cal J}_{n,2}}|\Zn(I)|\ \
\stackrel{d}{\leq} \ \ L+\sqrt{2 \log \frac{e(1+2\log^3\!n)|I_n|}{|I_n|/
\log^3\!n}} \ \ \leq \ \ L+4\sqrt{\log \log n}.
$$
One readily checks that $I \in {\cal J}_{n,2}$ implies 
$|I \cap I_n| \leq \sqrt{|I||I_n|/b_n}$. Thus (\ref{1a}) gives
\begin{equation*}
\begin{split}
\Pr_{f_n}\Bigl( \max_{I\in {\cal J}_{n,2}}|\Yn(I)| > \kappa_n \Br)
& \leq \Pr\Bigl( \max_{I\in {\cal J}_{n,2}}|\Zn(I)| +\frac{
\sqrt{2 \log n}-b_n}{\sqrt{b_n}} > \kappa_n \Br) \\
& \leq \Pr\Bl( L > \kappa_n -\frac{\sqrt{2 \log n} -b_n}{\sqrt{b_n}}
-4\sqrt{\log \log n} \Br)\\
& \ra 0\ \ \ \mbox{ since } \kappa_n =\sqrt{2 \log n} +O(1).
\end{split}
\end{equation*}

Finally, consider ${\cal J}_{n,3}:=\Bl\{I \in {\cal J}_n:\ I \cap I_n =
\emptyset\ $ or $\ |I| \leq |I_n|/\log^3\!n\ $ or $\ |I| > 
|I_n|\log^3\!n\Br\}$. Since
$I \in {\cal J}_{n,3}$ implies $|I \cap I_n| \leq \sqrt{|I||I_n|/
\log^3\!n}$, we get by (\ref{1a}),
$$
\Pr_{f_n}\Bigl( \max_{I\in {\cal J}_{n,3}}|\Yn(I)| > \kappa_n \Br)
\ \leq \ \Pr\Bigl( \max_{I\in {\cal J}_n}|\Zn(I)| > \kappa_n -
\frac{\sqrt{2 \log n}-b_n}{\sqrt{\log^3\!n}} \Br) 
\  \ra  \ \alpha,
$$
where the convergence follows from the following.

A sequence $\{c_n\}$ satisfies $\lim_n \Pr \Bl( \max_{I\in {\cal J}_n}
|\Zn(I)| > c_n\Br)=\alpha \ $ if and only if
\be  \label{kappa}
c_n \ =\ \sqrt{2\log n} + (2\log n)^{-1/2}\Bl(\frac{1}{2} \log \log n
+C(\alpha)\Br) +o\Bl((\log n)^{-1/2}\Br)
\ee
for a certain constant $C(\alpha)$. This follows from
Theorem~1.3 in Kabluchko~(2008) or, with some work, from the earlier
Theorem~1 in Siegmund and Venkatraman~(1995). 

Since ${\cal J}_n={\cal J}_{n,1} \cup {\cal J}_{n,2} \cup {\cal J}_{n,3}$
the theorem is proved. \hfill $\Box$
\bigskip

{\bf Proof of Theorem~\ref{ALR}:} We begin by showing that in the
null case of no signal, $\Yn=\Zn$, we have
\be  \label{ALRnull}
A_n\ =\ O_p(1).
\ee
Note that $A_n$ is an average of correlated random variables that do not possess
a finite first moment. In light of the converse to the strong law
it is thus not at all obvious that (\ref{ALRnull}) holds. For a proof
let $m>0$ and define the event
${\cal B}_{m,n}:= \{|\Zn((\frac{j}{n},\frac{k}{n}])| \leq C(\frac{
k-j}{n})+m \mbox{ for all }0\leq j<k \leq n\}$,
where $C(\del):=\sqrt{2 \log e/\del}$. Then Markov's inequality gives,
for $\lam >0$,
\begin{equation*}
\begin{split}
\Pr_0(A_n > \lam) &\leq \frac{1}{\lam n^2}  \sum_{j=0}^{n-1}
\sum_{k=j+1}^n \Ex \Bl[\exp \Bl(\frac{(\Zn( (\frac{j}{n},
\frac{k}{n}]))^2}{2}\Br) \ 1({\cal B}_{m,n})\Br]\ +\Pr({\cal B}_{m,n}^c)\\
&\leq \frac{1}{\lam n^2} \sum_{j=0}^{n-1} \sum_{k=j+1}^n
\int_{-C(\frac{k-j}{n})-m}^{C(\frac{k-j}{n})+m} e^{z^2/2} \frac{1}{
\sqrt{2\pi}} e^{-z^2/2} dz\ +\Pr({\cal B}_{m,n}^c)\\
&\leq \frac{1}{\lam n^2} \sum_{j=0}^{n-1} \sum_{k=j+1}^n
\Bl(\sqrt{2 \log \frac{en}{k-j}}+m\Br) +\Pr({\cal B}_{m,n}^c)\\
&\leq \frac{1}{\lam} \Bl(\frac{1}{n} \sum_{j=0}^{n-1} \int_0^1
\sqrt{2 \log e/u}\, du +m\Br) +\Pr({\cal B}_{m,n}^c)\\
& \leq \frac{4+m}{\lam} +\Pr(L >m)
\end{split}
\end{equation*}
by (\ref{B0}). This sum can be made arbitrarily small by choosing 
$m$ and $\lam$ appropriately, proving (\ref{ALRnull}).

To prove parts~1 and 3 together we consider $f_n$ with arbitrary
spatial extent and $|\mu_n|\sqrt{|I_n|}
 \geq \Bl(\sqrt{4 \log \frac{1}{|I_n|}}
+b_n\Br) /\sqrt{n}$ with $b_n \ra \infty$.
Set $\eps_n :=\min\Bl(1,b_n(\log \frac{e}{|I_n|})^{-1/2}
\Br)$ and ${\cal J}(I_n):=\{I \in {\cal J}_n: I \subset I_n$ and
$|I| \geq |I_n|(1-\eps_n/2)\}$. Then $\# {\cal J}(I_n) \geq
\lceil \eps_n n|I_n|/4\rceil^2$ since each of the $\lceil \eps_n 
n|I_n|/4\rceil$ smallest (largest)
design points in $cl(I_n)$ may serve as a left (right) endpoint
for some $I \in {\cal J}(I_n)$. Lemma~\ref{B1} gives
$$
\max_{I\in {\cal J}(I_n)}|\Zn(I)| \ \stackrel{d}{\leq}\ 
L+\sqrt{2\log \frac{e}{1-\eps_n/2}} \ \leq \ L+2.
$$
Together with $\frac{|I \cap I_n|}{\sqrt{|I||I_n|}} =
\sqrt{\frac{|I|}{|I_n|}} \geq \sqrt{1-\frac{\eps_n}{2}}$
and (\ref{1a}) we get
\bea
\min_{I \in {\cal J}(I_n)} |\Yn(I)| &\stackrel{d}{\geq} 
&\Bl(\sqrt{4\log \frac{1}{|I_n|}}+b_n\Br)
\sqrt{1-\frac{\eps_n}{2}} -L-2\\
& \geq & \sqrt{4 \log  \frac{1}{|I_n|}} +\frac{b_n}{9} -L-2,
\eea
since $(x+y)\sqrt{1 -\min(1,y)/2} \geq x+\frac{y}{9}\ $ for $x \in [0,2],
y>0$. Thus, writing $R_n:=\frac{b_n}{9} -L-2$,
\begin{equation*}
\begin{split}
A_n &\geq  \frac{\#{\cal J}(I_n)}{n^2} \min_{I \in {\cal J}(I_n)} 
\exp \Bl(\frac{(\Yn(I))^2}{2} \Br)\\
&\stackrel{d}{\geq} \frac{\eps_n^2 |I_n|^2}{4^2} \ \exp\Bl\{
2 \log  \frac{1}{|I_n|} +R_n\Bl(R_n/2 +\sqrt{4 \log  
\frac{1}{|I_n|}}\Br)\Br\}\\
&=\frac{\eps_n^2}{4^2}  \ \exp\Bl\{R_n\Bl(R_n/2 +\sqrt{4 \log  
\frac{1}{|I_n|}}\Br)\Br\}\\
&\geq \frac{\eps_n^2}{4^2} \Bl(\log  \frac{e}{|I_n|}\Br) \exp(R_n^2/2)
\ 1(R_n \geq 1)\\ 
&\stackrel{a.s.}{\ra} \infty \ \ \ \mbox{ since }R_n\stackrel{a.s.}{\ra} 
\infty \ \mbox{ and } \eps_n \sqrt{\log  \frac{e}{|I_n|}} \geq 1
\mbox{ eventually.}
\end{split}
\end{equation*}
The claim follows since $\tau_n =O(1)$ by (\ref{ALRnull}).
\medskip

For the proof of part~2 we partition ${\cal J}_n$
into ${\cal J}_{n,1}:=\{I \in {\cal J}_n: I \cap I_n \neq \emptyset$
 and $|I_n|/\log^4\!\frac{
1}{|I_n|} \leq |I| \leq |I_n|\log^4\!\frac{1}{|I_n|} \}$
and ${\cal J}_{n,2}:={\cal J}_n \cap {\cal J}_{n,1}^c$.
We show for $\JJ=\JJ_{n,1}, \JJ_{n,2}$ that
\be  \label{1thm2}
\frac{1}{n^2} \sum_{I \in \JJ} \Bl| \exp  \Bl(\frac{(\Yn(I))^2}{2} 
\Br) -\exp  \Bl(\frac{(\Zn(I))^2}{2}\Br) \Br| \ \stackrel{P}{\ra}\ 0.
\ee
Since it can be shown that in the null case $f_n\equiv 0$, the ALR
$A_n$ converges weakly to a continuous limit, the claim of part~2
follows from (\ref{1thm2}).

To prove (\ref{1thm2}) for $\JJ=\JJ_{n,1}$ we follow the proof
of part~2 of Theorem~\ref{scan} (set $b_n:=\log^4\!\frac{1}{|I_n|}$
there) and conclude $\max_{I\in {\cal J}_{n,1}}|\Zn(I)|
\stackrel{d}{\leq} L+2\sqrt{\log (3\log^4\!\frac{1}{|I_n|})}$.
Hence for a fixed constant $\lam$ which is specified below we obtain
$$
\Pr \Bl( {\cal A}_n\ :=\ \Bl\{\max_{I\in {\cal J}_{n,1}}|\Zn(I)|
\leq \lam \sqrt{\log \frac{1}{|I_n|}}\Br\}\Br) \ \ra \ 1.
$$
That proof also shows that every $I \in \JJ_{n,1}$ is contained
in a certain interval of length $(1+2\log^4\!\frac{1}{|I_n|})|I_n|$,
thus $\# \JJ_{n,1} \leq \Bl((1+2\log^4\!\frac{1}{|I_n|})
n|I_n|\Br)^2$. On the event ${\cal A}_n$ we have, by (\ref{1a}),
\bea
\lefteqn{\frac{1}{n^2} \sum_{I \in \JJ_{n,1}} \Bl| \exp  
\Bl(\frac{(\Yn(I))^2}{2}\Br)-\exp  \Bl(\frac{(\Zn(I))^2}{2}\Br)\Br|}\\
&\leq & \frac{\# \JJ_{n,1}}{n^2}\ 2\ \exp \Bl(\max_{I\in {\cal J}_{n,1}}
\frac{(\Zn(I))^2}{2} +\frac{n|\mu_n|^2|I_n|}{2}
+\max_{I\in {\cal J}_{n,1}}|\Zn(I)|\sqrt{n|I_n|}|\mu_n| \Br)\\
&\leq & 18 \Bl(\log\!\frac{1}{|I_n|}\Br)^8\ |I_n|^2 \ \exp \Bl(\frac{
\lam^2 \log\frac{1}{|I_n|}}{2} +\frac{K^2}{2} \log\frac{1}{|I_n|}
+\lam K \log\frac{1}{|I_n|}\Br)\\
&=& 18 \Bl(\log\!\frac{1}{|I_n|}\Br)^8\ |I_n|^{2-\frac{K^2}{2}-
\frac{\lam^2}{2}-\lam K}.
\eea
Since $\frac{K^2}{2}<2$ we can choose $\lam=\lam(K)>0$ such
that the above expression goes to 0 as $|I_n|\ra 0$, proving
(\ref{1thm2}) for $\JJ=\JJ_{n,1}$.

To prove (\ref{1thm2}) for $\JJ=\JJ_{n,2}$, we proceed similarly
as in the proof of (\ref{ALRnull}) and employ the event 
${\cal B}_{m,n}$ defined there. Then for $\lam,m>0$, Markov's
inequality gives
\bea
\lefteqn{ \Pr_{f_n} \Bl(\frac{1}{n^2} \sum_{I \in \JJ_{n,2}} \Bl|\exp
\Bl(\frac{(\Yn(I))^2}{2}\Br)-\exp  \Bl(\frac{(\Zn(I))^2}{2}\Br)\Br|
\ >\ \lam \Br)}\\
&\leq & \frac{1}{\lam n^2} \sum_{I \in \JJ_{n,2}} \Ex_{f_n} \Bl(\Bl| 
\exp \Bl(\frac{(\Yn(I))^2}{2}\Br)-\exp  \Bl(\frac{(\Zn(I))^2}{2}\Br)
\Br|1({\cal B}_{m,n})\Br) +\Pr({\cal B}_{m,n}^c).
\eea
Since (\ref{B0}) gives $\Pr({\cal B}_{m,n}^c) \leq \Pr(L>m) \ra 0$
as $m \ra \infty$, it is enough to show that for any fixed $m$ the 
above expectation converges to 0 as $n \ra \infty$, uniformly in
$I \in \JJ_{n,2}$. Using $\Zn(I) \sim $ N(0,1) and writing
$\del_I:=\sqrt{n}|\mu_n| \frac{|I \cap I_n|}{\sqrt{|I|}}$
and $C(|I|):=\sqrt{2\log e/|I|}$, we get with (\ref{1a}),
\bean
\lefteqn{\Ex_{f_n} \Bl(\Bl| \exp \Bl(\frac{(\Yn(I))^2}{2}\Br)-
\exp  \Bl(\frac{(\Zn(I))^2}{2}\Br)\Br|1({\cal B}_{m,n})\Br)}\nn \\
&\leq & \frac{1}{\sqrt{2\pi}} \int_{-C(|I|)-m}^{C(|I|)+m} \Bl|
\exp(z \del_I +\del_I^2/2) -1\Br| dz \nn\\
& \leq & \sqrt{\frac{2}{\pi}} \exp \Bl( \del_I(C(|I|)+m)+\del_I^2/2
\Br)\Bl(\del_I (C(|I|)+m)^2 +\del_I^2(C(|I|)+m)/2\Bl) \label{2thm2}
\eean
by bounding the function $z \mapsto \exp(z \del +\del^2/2)-1$
above and below by the Mean Value Theorem. Next we show
that, as $n \ra \infty$,
\be \label{3thm2}
\del_I \ra 0 \ \ \ \mbox{ and }\ \ \ \del_I C^2(|I|) \ra 0\ \ \ 
\mbox{ uniformly in }I \in \JJ_{n,2}.
\ee
This conclusion then also holds for the expression in (\ref{2thm2}),
and (\ref{1thm2}) follows.

To prove (\ref{3thm2}), note that $\del_I=K\sqrt{\log \frac{1}{|I_n|}}
\ \frac{|I \cap I_n|}{\sqrt{|I||I_n|}}$. 
If $I \cap I_n =\emptyset$, then $\del_I=0$.
If $|I| <|I_n|/\log^4\!\frac{1}{|I_n|}$, then the bound $|I \cap I_n|
\leq |I|$ yields $\del_I \leq K(\log \frac{1}{|I_n|})^{-3/2}$,
while the monotonicity of the function  $x \mapsto \sqrt{x} \log e/x$
for $ x\in (0,e^{-1})$ gives
\bea
\del_I C^2(|I|) & \leq & 2K \sqrt{\log \frac{1}{|I_n|}} \sqrt{\frac{|I|
}{|I_n|}} \log \frac{e}{|I|} \\
& \leq & 2K \Bl(\log \frac{1}{|I_n|}\Br)^{-3/2} 
\log \frac{e \log^4\!\frac{1}{|I_n|}}{|I_n|}\\
& \leq & 6K \Bl(\log \frac{1}{|I_n|} \Br)^{-1/2}
\eea
if $n$ is large enough so that $|I_n|/\log^4\!\frac{1}{|I_n|} \leq
e^{-1}$.

If $|I| >|I_n|\log^4\!\frac{1}{|I_n|}$, then the bound $|I \cap I_n|
\leq |I_n|$ yields again $\del_I \leq K(\log \frac{1}{|I_n|})^{-3/2}$,
while $\del_I C^2(|I|) \leq 2K(\log \frac{1}{|I_n|})^{-3/2}
\log \frac{e}{|I|} \leq 2K(\log \frac{1}{|I_n|})^{-1/2}$.
\hfill $\Box$
\bigskip

{\bf Proof of Theorem~\ref{ALRapprox}:}
Before proceeding to the proof, we sketch an explanation for
the choice of the grid spacing $d_{\ell}$. For given $I_n$,
let $\ell$ be such that the intervals in $\Iapp (\ell)$
have length about $|I_n|$. Thus $m_{\ell} \sim n|I_n|$,
i.e. $\ell \approx \log_2 \frac{e}{|I_n|}$. An interval $I$
results in a significant likelihood ratio provided its
endpoints lie in a $\frac{\eps_n}{2} |I_n|$ neighborhood of 
the endpoints
of $I_n$, where $\eps_n :=\min\Bl(1,b_n(\log \frac{e}{
|I_n|})^{-1/2}\Br)$. Thus the number of significant intervals
in $\Iapp (\ell)$ is  $\sim (\eps_n |I_n|n/d_{\ell})^2$. Under
(\ref{optcondition}) the size
of the corresponding likelihood ratios is $\gg \frac{
(\log_2 \frac{e}{|I_n|})^p}{|I_n|}$ for arbitrary $p>0$. Thus
optimality of $A_{n,cond}$ obtains if the number
of significant intervals in $\Iapp (\ell)$ is at least
$ \# \Iapp \frac{|I_n|}{(\log_2 \frac{e}{|I_n|})^p}$
for some fixed $p>0$, and even if the number is smaller by
some factor $\eps_n^k$, since $\eps_n \sqrt{\log_2 \frac{e}{|I_n|}}
\geq 1$. Solving this inequality for $d_{\ell}$
yields $d_{\ell} \leq \sqrt{\frac{n^2 |I_n|\ell^p}{\#\Iapp}}$. 
Requiring $\# \Iapp \sim n (\log n)^q$ for some $q>0$ for
computational efficiency suggests the choice $d_{\ell} \sim
\sqrt{m_{\ell} \ell^p (\log n)^{-q}}$. Computing
$\# \Iapp$ shows that this choice of $d_{\ell}$ is indeed
consistent with $\# \Iapp \sim n (\log n)^q$ provided $p>1$.
Further, it will be seen that $A_{n,cond}=O_p(1)$ under the null 
hypothesis requires
$\sum_{\ell} \ell^{1/2-p} < \infty$, i.e. $p>3/2$. Finally, 
optimal detection for very small intervals $I_n$ requires
that their endpoints are approximated exactly, i.e. it is
necessary to have $d_{\ell}=1$
for large $\ell$. Thus we need an appropriate combination
of a small $p>3/2$ and a large $q$. Since the required large $q$
results in a noticeably worse computation time, we prefer to
stick to a $O(n \log^2 n)$ algorithm by setting $p=8/5$, $q=2$,
and by explicitly considering all small intervals containing up
to $\log n$ design points in lieu of choosing a larger $q$.

We now prove the theorem, starting with the 
claim about the computational complexity. Since
$\Iapp (\ell)$ allows only every $d_{\ell}$th design point
as a potential endpoint for an interval, there are at most $\lceil
n/d_{\ell} \rceil$ left endpoints. For each left endpoint there
are at most $\lceil 2m_{\ell}/d_{\ell} \rceil$ right endpoints
since each interval contains not more than $2m_{\ell}$ design points.
Thus
\begin{gather}
\# \Iapp (\ell) \ \leq \ \Bl\lceil \frac{n}{d_{\ell}} \Br\rceil
  \Bl\lceil \frac{2m_{\ell}}{d_{\ell}} \Br\rceil \ \leq \ 3 
  \frac{n \log^2\!n}{\ell^{8/5}}\ \ \ \mbox{ and}  \label{3a} \\
\# \Ismall \ \leq \ n \log n \ \ \ \ \mbox{ since } m_{\ell_{max}} 
  \leq \log n. \mbox{ Hence} \label{3aa} \\
\# \Iapp \ \leq \ \sum_{\ell=1}^{\ell_{max}} 3 \frac{n \log^2\!n}{
  \ell^{8/5}} + n \log n \ \leq \ 9 n \log^2\!n. \label{3b} 
\end{gather}

\nin $\exp \Bl(\frac{(\Yn(I))^2}{2}\Br)$ can be evaluated in a constant
number of steps after an initial one-time computation of the 
cumulative sum vector of $(Y_i, 1 \leq i \leq n)$. Since that
computation has complexity $O(n)$, the overall computational
complexity of computing $A_{n,cond}$ is dominated by the cardinality of 
$\Iapp$ and hence is $O( n \log^2\!n)$.

Next we show that for $f_n \equiv 0$:
\be  \label{3c}
A_{n,cond}\ =\ O_p(1)\ \ \mbox{ as } n \ra \infty.
\ee
Proceeding as in the proof of (\ref{ALRnull}) it is enough to
show that
\be  \label{3d}
\frac{1}{\# \Iapp} \sum_{I \in \Iapp} \sqrt{ 2\log \frac{1}{|I|}}
\ =\ O(1).
\ee
Since $I \in \Iapp (\ell)$ implies $|I| > \frac{m_{\ell}}{n}
= 2^{-\ell}$, we obtain with (\ref{3a}) and (\ref{3aa}),
\bea
\sum_{I \in \Iapp} \sqrt{ 2\log \frac{1}{|I|}}
& \leq & \sum_{\ell=1}^{\ell_{max}} \#\Iapp (\ell) \sqrt{2\ell}\ +
n (\log n) \sqrt{2 \log n}\\
& \leq & 5 n \log^2\!n \sum_{\ell=1}^{\infty} \frac{\ell^{1/2}}{
\ell^{8/5}}\ +n (\log n)^2\\
& \leq & 56 n \log^2\!n.
\eea
On the other hand, $\#  \Iapp \geq \#  \Iapp(2) \geq \frac{1}{10}
n (\log n)^2$ by considerations similar to those establishing
(\ref{3a}). (\ref{3d}) follows.

To establish optimality of $A_{n,cond}$, we proceed as in the proof
of Theorem~\ref{ALR} and consider $f_n$ with arbitrary spatial extent
and $|\mu_n| \sqrt{|I_n|} \geq \Bl(\sqrt{2 \log \frac{1}{|I_n|}}
+b_n\Br) /\sqrt{n}$, where $b_n \ra \infty$. As before we define
$\eps_n :=\min\Bl(1,b_n(\log \frac{e}{|I_n|})^{-1/2}\Br)$ and
${\cal J}(I_n):=\{I \in {\cal J}_n: I \subset I_n$ and
$|I| \geq |I_n|(1-\eps_n/2)\}$. Intervals $I \in {\cal J}(I_n)$ 
contribute significant LR statistics to $A_{n,cond}$. Since we now
require that the endpoints of these intervals fall on a $d_{\ell}$-
grid, there are now many fewer of these intervals. But this is more
than compensated by the small cardinality of $\Iapp$ appearing in 
the divisor of $A_{n,cond}$. This fact allows us to detect $f_n$ with
a norm that is smaller than in Theorem~\ref{ALR}. In more detail,
take the integer $\ell$ so
$m_{\ell} <n|I_n|(1-\eps_n/4) \leq 2m_{\ell}$.
\begin{Lemma}   \label{numint}
\bea
\mbox{If }\ell &\leq & \ell_{max}, \mbox{ then } \frac{\# \Bl(
{\cal J}(I_n) \cap \Iapp (\ell) \Br)}{\# \Iapp} 
\geq \frac{\eps_n^2 |I_n|}{9^3 \Bl(\log_2 \frac{e}{|I_n|}\Br)^{8/5}}.\\
\mbox{If }\ell & >&  \ell_{max}, \mbox{ then } \frac{\# \Bl(
{\cal J}(I_n)\cap \Ismall \Br)}{\# \Iapp} \geq 
\frac{\eps_n^2 |I_n|}{32^2 \Bl(\log_2 \frac{e}{|I_n|}\Br)^2}.
\eea
\end{Lemma}
As in the proof of Theorem~\ref{ALR} we find $\min_{I \in {\cal J}(I_n)}
|\Yn(I)| \stackrel{d}{\geq} \sqrt{2 \log  \frac{1}{|I_n|}}
+R_n$, where $R_n:=\frac{b_n}{9} -L-2$. Thus in the case $\ell \leq  
\ell_{max}$, Lemma~\ref{numint} gives
\begin{equation*}
\begin{split}
A_{n,cond} &\geq  \frac{\# \Bl({\cal J}(I_n)\cap \Iapp (\ell)\Br)}{
\# \Iapp} \min_{I \in {\cal J}(I_n)}
\exp \Bl(\frac{(\Yn(I))^2}{2} \Br) \\
&\stackrel{d}{\geq} \frac{\eps_n^2 |I_n|}{9^3 \Bl(\log_2 \frac{e}{|I_n|}
\Br)^{8/5}} \ \exp\Bl\{\log  \frac{1}{|I_n|} +R_n\Bl(R_n/2 +\sqrt{2 \log
\frac{1}{|I_n|}}\Br)\Br\}\\
&= \frac{\eps_n^2}{9^3 \Bl(\log_2 \frac{e}{|I_n|}\Br)^{8/5}}
\ \exp\Bl\{R_n\Bl(R_n/2 +\sqrt{2 \log \frac{1}{|I_n|}}\Br)\Br\}\\
&\geq C\eps_n^2 \Bl(\log  \frac{e}{|I_n|}\Br) \exp(R_n^2/2)
\ 1(R_n \geq 1)\ \mbox{ for some universal }C>0\\
&\stackrel{a.s.}{\ra} \infty \ \ \ \mbox{ since }R_n\stackrel{a.s.}{\ra}
\infty \ \mbox{ and } \eps_n \sqrt{\log  \frac{e}{|I_n|}} \geq 1
\mbox{ eventually.}
\end{split}
\end{equation*}
In the case $\ell > \ell_{max}$, the same conclusion obtains by
using $\Ismall$ in place of $\Iapp (\ell)$.
The claim then follows since the critical value of $A_{n,cond}$
stays bounded, by (\ref{3c}).

Thus the crucial difference with Theorem~\ref{ALR} is the stronger
inequality provided by Lemma~\ref{numint}. The corresponding
inequality for Theorem~\ref{ALR} is $\# {\cal J}(I_n)/n^2 \geq
\eps_n^2 |I_n|^2/4^2$, and the extra term $|I_n|$ causes the loss
of efficiency in the case when the spatial extent $|I_n|$ is
small.

It remains to prove Lemma~\ref{numint}. Elementary considerations
show that one can find sets ${\cal I}_{n,left}$ and ${\cal I}_{n,right}$,
each consisting of $p:=\lceil \eps_n n|I_n|/8 \rceil$ consecutive 
integers, such that $(j,k) \in {\cal I}_{n,left} \times 
{\cal I}_{n,right}$ implies $(\frac{j}{n},\frac{k}{n}] \in 
{\cal J}(I_n)$, and also $m_{\ell} <k-j \leq 2 m_{\ell}$ if
$\ell \leq \ell_{max}$, resp. $k-j \leq m_{\ell_{max}}$ if 
$\ell > \ell_{max}$. Thus, in the latter case, we immediately obtain
$\# \Bl({\cal J}(I_n)\cap \Ismall \Br) \geq p^2$, and the claim of the
Lemma obtains with (\ref{3b}),
$n|I_n|\geq 1$, and $\log n \leq \frac{4}{3}\log_2  \frac{e}{|I_n|}$,
which follows from $\ell > \ell_{max}$.
In the case
$\ell \leq \ell_{max}$, only  a subset of the $p^2$ intervals 
$\Bl\{(\frac{j}{n},\frac{k}{n}]$: $(j,k) \in {\cal I}_{n,left} \times
{\cal I}_{n,right}\Br\}$ belongs to ${\cal J}(I_n)
\cap \Iapp (\ell)$, namely those for which both $j$ and $k$ lie
on the $d_{\ell}$-grid. The number of such indices $j$ is at least
$\lfloor \frac{p}{d_{\ell}}\rfloor \geq \lceil \frac{ \eps_n
\sqrt{ n|I_n|} \log n}{9 \ \ell^{4/5}} \rceil$ for $n$ large enough,
where this inequality follows from the fact that 
$\ell \leq \ell_{max}$ implies
$n|I_n| > m_{\ell} \geq (\log n)/2$ and $d_{\ell}\geq (\log n)^{3/10}$;
further, $\eps_n \sqrt{\log n} \ra \infty$ by the definition of $\eps_n$.
The same bound obtains for the number of indices $k$ that lie on the
$d_{\ell}$-grid. The claim of the Lemma then follows with (\ref{3b})
and $\ell \leq \log_2  \frac{e}{|I_n|}$, which is a consequence of
the definition of $\ell$. \hfill $\Box$
\bigskip

\subsection*{References}

\begin{description}
\item[] Arias-Castro, E., Donoho, D.L., and Huo, X. (2005).
Near-optimal detection of geometric objects by fast multiscale
methods. \ {\sl IEEE Trans. Inform. Th.\ \textbf{51}}, 2402--2425.
\item[] Burnashev, M.V. and Begmatov, I.A.~(1990).
On a problem of detecting a signal that leads to stable distributions.\
{\sl Theory Probab. Appl.\ \textbf{35}}, 556--560.
\item[] Cai, T., Jeng, X.J., and Li, H. (2011).
Robust detection and identification of sparse segments in ultra-high
dimensional data analysis. Manuscript.
\item[] Chan, H.P. (2009). Detection of spatial clustering with
average likelihood ratio test statistics. \ {\sl Ann. Statist.\ 
\textbf{37}}, 3985--4010.
\item[] Chan, H.P. and Walther, G. (2011). Detection with the
scan and the average likelihood ratio. \ Manuscript. January 2011.
\item[] Chan, H.P. and Zhang, N.R. (2009). Local average likelihood
ratio test statistics with applications in genomics and change-point
detection. \ Manuscript.
\item[] D\"{u}mbgen, L. (1998).
New goodness-of-fit tests and their application to nonparametric 
confidence sets.\
{\sl Ann.\ Statist. \textbf{26}}, 288--314. 
\item[] D\"{u}mbgen, L. and Spokoiny, V.G. (2001).
        Multiscale testing of qualitative hypotheses. \
        {\sl Ann.\ Statist.\ \textbf{29}}, 124--152.
\item[] D\"{u}mbgen, L. and Walther, G. (2008).
 Multiscale inference about a density. \
{\sl Ann.\ Statist. \textbf{36}}, 1758--1785.
\item[] Gangnon, R.E. and Clayton, M.K. (2001).
The weighted average likelihood ratio test for spatial disease
clustering.\ {\sl Statistics in Medicine \textbf{20}}, 2977--2987.
\item[]  Glaz, J. and Balakrishnan, N. (ed.) (1999).
{\sl Scan statistics and Applications.} \
Birkh\"{a}user, Boston
\item[]  Glaz, J., Naus, J., and Wallenstein, S. (2001).
{\sl Scan Statistics.} \ Springer, New York
\item[]  Glaz, J., Poznyakov, V., and Wallenstein, S. (ed.) (2009). 
{\sl Scan Statistics: Methods and Applications.} \ Birkhauser, Boston
\item[] Kabluchko, Z. (2008). Extreme-value analysis of
standardized Gaussian increments. Manuscript.
\item[] Kulldorff, M. (1997). A spatial scan statistic. \
{\sl Commun. Statist. - Theory Meth. \textbf{26}}, 1481--1496.
\item[] Lepski, O.V. and Tsybakov, A.B. (2000).
Asymptotically exact nonparametric hypothesis testing in sup-norm
and at a fixed point. \
{\sl Probab. Theory Related Fields \textbf{117}}, 17--48.
\item[] Neill, D. and Moore, A. (2004).
A fast multi-resolution method for detection of significant spatial
disease clusters.
{\sl Adv. Neur. Info. Proc. Sys. \textbf{10}}, 651--658.
\item[] Neill, D. (2009). An empirical comparison of spatial
scan statistics for outbreak detection.\ {\sl Internat. Journal of
Health Geographics \textbf{8}}, 1--16.
\item[] Rohde, A. (2008). Adaptive goodness-of-fit tests based
on signed ranks.\
{\sl Ann.\ Statist.\ \textbf{36}}, 1346--1374.
\item[] Rufibach, K. and Walther, G. (2010). The block criterion 
for multiscale inference about a density, 
with applications to other multiscale problems. \
{\sl Journal of Computational and Graphical Statistics \textbf{19}}, 
175--190.
\item[] Shiryaev, A.N. (1963).
On optimum methods in quickest detection problems. \
{\sl Theory Probab. Appl. \textbf{8}}, 22-46.
\item[] Siegmund, D. and Venkatraman, E.S. (1995).
Using the generalized likelihood ratio statistic for sequential
detection of a change-point. \
{\sl Ann.\ Statist. \textbf{23}}, 255--271.
\item[] Siegmund, D. (2001). Is peak height sufficient?\
{\sl Genetic Epidemiology \textbf{20}}, 403--408.
\item[] Walther, G. (2010). Optimal and fast detection of
spatial clusters with scan statistics. \ 
{\sl Ann.\ Statist.\ \textbf{38}}, 1010-1033.
\end{description}

\end{document}